\documentclass[traditabstract]{aa} %
\usepackage{txfonts}
\usepackage{graphicx}
\usepackage{natbib}

\newcommand{\hwvir}{\mbox{\object{HW\,Virginis}}}
\newcommand{\hw}{\mbox{\object{HW\,Vir}}}
\newcommand{\huaqr}{\mbox{\object{HU\,Aqr}}}
\newcommand{\nnser}{\mbox{\object{NN\,Ser}}}
\newcommand{\dpleo}{\mbox{\object{DP\,Leo}}}
\newcommand{\rrcae}{\mbox{\object{RR\,Cae}}}
\newcommand{\nsvs}{\mbox{\object{NSVS\,14256825}}}
\newcommand{\hs}{\mbox{\object{HS\,0705+67}}}

\newcommand{\ten}[2]{#1\times 10^{#2}}
\newcommand{\oc}{$O\!-\!C$}
\newcommand{\oclin}{$O\!-\!C_\mathrm{lin}$}
\newcommand{\oclinm}{$O\!-\!C_\mathrm{lin,1}$}
\newcommand{\oclinl}{$O\!-\!C_\mathrm{lin,2}$}

\newcommand{\rsun}{$R_\odot$}
\newcommand{\msun}{$M_\odot$}
\newcommand{\mjup}{$M_\mathrm{Jup}$}
\newcommand{\chisq}{$\,\chi^2$}

\begin{document}

\title{The quest for companions to post-common envelope binaries}
\subtitle{III. A reexamination of \hwvir}

\author{
K.~Beuermann \inst{1} \and 
S.~Dreizler \inst{1} \and 
F.~V.~Hessman \inst{1} \and 
J.~Deller \inst{1}  
} 

\institute{
Institut f\"ur Astrophysik, Georg-August-Universit\"at, Friedrich-Hund-Platz 1, D-37077 G\"ottingen, Germany 
}
\date{Received 12 April 2012; accepted 11 June 2012}

\authorrunning{K. Beuermann et al.} 
\titlerunning{The quest for companions to post-common envelope binaries III}

\abstract{\vspace*{-2mm} We report new mid-eclipse times of the short-period sdB/dM
  binary \hwvir, which differ substantially from the times predicted
  by a previous model. The proposed orbits of the two planets in that
  model are found to be unstable. We present a new secularly stable
  solution, which involves two companions orbiting \hw\ with periods
  of 12.7\,yr and $55\pm 15$\,yr.  For orbits coplanar with the binary,
  the inner companion is a giant planet with mass
  $M_3\,\mathrm{sin}\,i_3\!\simeq\!14$\,\mjup\ and the outer one a
  brown dwarf or low-mass star with a mass of
  $M_4\,\mathrm{sin}\,i_4\!=\!30-120$\,\mjup. Using the {\tt mercury6}
  code, we find that such a system would be stable over more than
  $10^7$\,yr, in spite of the sizeable interaction. Our model fits the
  observed eclipse-time variations by the light-travel time effect
  alone, without invoking any additional process, and provides support
  for the planetary hypothesis of the eclipse-time variations in close
  binaries. The signature of non-Keplerian orbits may be visible in
  the data.}


\maketitle


\section{Introduction}

\vspace*{-1mm}
Periodic or quasiperiodic variations of the mid-eclipse times in close
binaries have been observed for decades and variously ascribed to
activity cycles of the secondary star \citep{applegate92}, apsidal
motion \citep{todoran72}, or the response to a third body orbiting the
binary \citep{natherrobinson74}. More recently, \citet{leeetal09}
assigned the complex $O\!-\!C$ (observed minus calculated)
eclipse-time variations in the detached sdB/dM binary \hw\ with a
2.8-h orbital period to the light-travel time (LTT) effect caused by
the orbital motion of two giant planets. Thereafter, an increasing
number of eclipsing post-common envelope binaries (PCEB), both with
sdB and white-dwarf primaries, were proposed as having eclipse-time
variations possibly due to planetary or brown-dwarf
companions. Systems with one proposed companion include HS0705+67,
\dpleo, HS2231+24, and NSVS14256825
\citep{qian-hs0705,qian-hs2231,qian-dpleo,beuermannetal11,beuermannetal12}.
Further PCEB that may harbor more than one companion are \nnser\
\citep{beuermannetal10}, UZ~For \citep{potteretal11}, \rrcae\ and
NY~Vir \citep{qian-rrcae,qian-nyvir}, QS~Vir \citep{parsonsetal10},
and HU~Aqr \citep{qian-huaqr,hinseetal12}.

The question of secular stability does not arise for binaries with a
single planet, but must be considered for the proposed systems of
companions. Such analyses were not included in most original
publications, with the exception of the two planets that orbit \nnser,
which represent a stable resonant pair \citep{beuermannetal10}. The
system suggested by \citet{qian-huaqr} for \huaqr, on the other hand,
was shown to be secularly unstable
\citep{horneretal11,wittenmyeretal11}, although a different stable
solution may exist \citep{hinseetal12}. To be sure, the planetary
hypothesis of the observed eclipse time variations has met with some
scepticism \citep[e.g.][]{wittenmyeretal11}, but a sufficiently
well-defined alternative is not in sight.

A more general critique of the models of \citet{leeetal09} and
\citet{qian-huaqr} relates to their combining the LTT effect with a
long-term period decrease on a time scale of $\tau\!=\!P/\dot
P\!\simeq\!10^7$\,yr, supposedly produced by some other mechanism.
Gravitational radiation leads to a period decrease, but is entirely
negligible with $\tau\!\simeq\!\ten{3}{9}$\,yr. Magnetic braking is a
more adequate contender for a long-term period decrease, but cannot
explain the recent period increase in \hw. Applegate's (1992)
mechanism allows variations of both signs, but most authors agree that
it is too feeble to produce the large observed eclipse-time variations
\citep[e.g.][]{brinkworthetal06,watsonmarsh10}. If, however, a so far
unknown super-Applegate effect were operative in close binaries, it
could conceivably account for the entire observed effect and render
the arbitrary division into variations produced by different
mechanisms meaningless. Indeed, it is the spectre of a super-Applegate
mechanism that has historically prevented us from considering in close
binary systems what, for unevolved single stars, has become the
default situation: complex planetary systems \citep{lovisetal11}. The
discovery by KEPLER of several circumbinary planetary systems in
non-evolved eclipsing binaries \citep{kepler16,kepler34b35b} shows us
that it is not unreasonable to consider the same for evolved binaries,
although their planetary systems possibly do not not survive the
common-envelope phase unscathed.  In this light, the proposed circumbinary
planetary systems detected by the LTT method must be carefully
re-scrutinized.

Here, we present new mid-eclipse times for \hw\ that deviate
significantly from the Lee et al. (2009) prediction. We show that the
planetary system proposed by Lee et al. is secularly unstable and,
therefore, untenable. We find that a secularly stable two-companion
model can be devised, in which the observed eclipse-time variations
are due to the LTT effect alone.

\section{The data base}

\begin{table}[t]
\begin{flushleft}
\caption{New primary mid-eclipse times $T_\mathrm{ecl}$ of \hw\
    and \oclinm\ residuals for the MONET/N data relative to the
    ephemeris of Eq.~\ref{eq:monet}.}
\begin{tabular}{ccccl}
\hline \\[-1ex]
Cycle   &  BJD(TT)      & Error    &  $O\!-\!C_\mathrm{lin,1}$  & Origin\\
            & 2400000+       & (days)   &  (days) & \\[0.5ex]
\hline\\[-1ex]
 75123    &    54498.877674 & 0.000014 &                    N.A.      & AAVSO  \\
 75890    &    54588.401364 & 0.000040 &                    N.A.      & AAVSO  \\
 76004    &    54601.707367 & 0.000016 &                    N.A.      & AAVSO \\
 76063    &    54608.593786 & 0.000019 &                    N.A.      & AAVSO \\
 76089    &    54611.628553 & 0.000012 &                    N.A.      & AAVSO \\
 78062    &    54841.916149 & 0.000010 &                    N.A.      & AAVSO \\
 84077    &    55543.984048 & 0.000008 &  \hspace{-2.0mm}$-$0.000007  & MONET/N \\ 
 84120    &    55549.003005 & 0.000008 &                    0.000009  & MONET/N \\
 84180    &    55556.006176 & 0.000009 &                    0.000006  & MONET/N \\ 
 84411    &    55582.968393 & 0.000009 &                    0.000006  & MONET/N  \\
 84428    &    55584.952622 & 0.000010 &                    0.000003  & MONET/N  \\
 84488    &    55591.955807 & 0.000010 &                    0.000014  & MONET/N \\
 84497    &    55593.006274 & 0.000007 &                    0.000005  & MONET/N \\
 84600    &    55605.028372 & 0.000007 &  \hspace{-2.0mm}$-$0.000010  & MONET/N \\
 84608    &    55605.962117 & 0.000015 &  \hspace{-2.0mm}$-$0.000022  & MONET/N \\
 84693    &    55615.883298 & 0.000007 &  \hspace{-2.0mm}$-$0.000004  & MONET/N \\
 84863    &    55635.725619 & 0.000006 &  \hspace{-2.0mm}$-$0.000007  & MONET/N \\
 84967    &    55647.864460 & 0.000007 &                    0.000001  & MONET/N \\
 84976    &    55648.914932 & 0.000007 &  \hspace{-2.0mm}$-$0.000004  & MONET/N \\
 85026    &    55654.750921 & 0.000006 &                    0.000007  & MONET/N\\
 85249    &    55680.779371 & 0.000007 &  \hspace{-2.0mm}$-$0.000003  & MONET/N \\
 85266    &    55682.763597 & 0.000015 &  \hspace{-2.0mm}$-$0.000009  & MONET/N \\
 87093    &    55896.010239 & 0.000007 &                    0.000005  & MONET/N \\
 87589    &    55953.903110 & 0.000018 &  \hspace{-2.0mm}$-$0.000023  & MONET/N \\
 87624    &    55957.988315 & 0.000007 &  \hspace{-2.0mm}$-$0.000003  & MONET/N \\
 87787    &    55977.013609 & 0.000007 &                    0.000004  & MONET/N \\[0.5ex]
\hline\\[-2ex]                                                              
\end{tabular}
\label{tab:monetaavso}
\end{flushleft}
\end{table}

Starting in December 2010, we monitored the V=10.9 mag binary \hw\
with the MONET/North telescope at the University of Texas' McDonald
Observatory via the MONET browser-based remote-observing
interface. The photometric data were taken with an Apogee ALTA E47+
1k$\times$1k CCD camera mostly in the $I_\mathrm{c}$-band with
exposure times of 10 or 20\,s. Since there is no suitable comparison
star in our $5'\times5'$ field, we included only observations obtained
under photometric conditions. The light curves were analyzed using the
heuristic mathematical model described in paper II of this series
\citep{beuermannetal12}. Table~\ref{tab:monetaavso} lists the 20 new
primary mid-eclipse times along with their formal 1-$\sigma$ errors,
which range from 0.5 to 1.6\,s. Fitting the MONET/N mid-eclipse times
alone yields the linear ephemeris valid in 2010/2012
\begin{equation}
T_\mathrm{ecl} = \mathrm{BJD(TT)}\,2455543.984055(2) + 0.116719555(2)\,E.
\label{eq:monet}
\end{equation}
The residuals from Eq.~\ref{eq:monet} are listed in
Table~\ref{tab:monetaavso} as \oclinm. Their rms value of 0.8\,s is
consistent with the errors obtained from the formal fits to the light
curves. Because of the effects of the additional bodies in the system,
the period of Eq.~\ref{eq:monet} is not necessarily identical to the
binary period.

A large body of primary and secondary mid-eclipse times of \hw\ is
available in the literature. The SAAO group monitored \hw\ between
1984 and 2002 \citep{marangkilkenny89} and
\citet{kilkennyetal91,kilkennyetal94,kilkennyetal00,kilkennyetal03},
reporting a total of 111 primary mid-eclipse times with errors mostly
as small as 2\,s.  The measurements of \citet{leeetal09} overlap with
the SAAO data and extend the coverage to 2009, with a gap in
2004. Additional mid-eclipse times were published by \citet{woodetal93},
\citet{gurolselan94}, \citet{woodsaffer99},
\citet{cakirlidevlen99}, \citet{kissetal00}, Agerer \& H\"ubscher
(2000, 2002, 2003), \citet{ibanogluetal04}, and
\citet{bratetal08,bratetal09,bratetal11}. Further timings were
drawn from the Japanese VSNET archive
\citep{vsnet}\footnote{http://www.kusastro.kyoto-u.ac.jp/vsnet/}, and
unanalyzed light curves were obtained from the AAVSO archive
\citep{aavso}\footnote{http://www.aavso.org/ql}. We determined
mid-eclipse times for the AAVSO data, using the same fitting method as
for the MONET data. The resulting new primary mid-eclipse times are
included in Table~\ref{tab:monetaavso}.

\begin{figure}[t]
\includegraphics[height=89mm,angle=-90,clip]{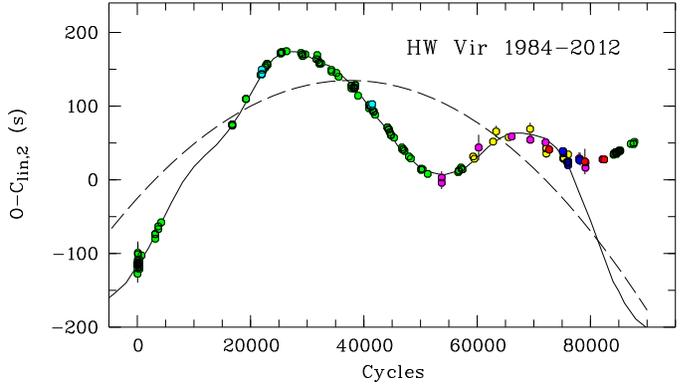}
\caption[chart]{$O-C_\mathrm{lin,2}$ residuals of the mid-eclipse times
  from the linear ephemeris used by \citet{leeetal09} along with their
  model curves for the two-companion model (solid) and the underlying
  quadratic ephemeris (dashed). The data are from SAAO
  (green), Wood et al (cyan blue), Lee et al. (yellow), BAV and VSNET
  (magenta), AAVSO (blue), BRNO (red), and MONET/North (green). }
\label{fig:data}
\end{figure}

The general picture that emerges from the entire body of eclipse times
collected between 1984 and 2012 is that of a smooth long-term
\oc\ variation of complex shape
\citep[e.g.][]{kilkennyetal03,ibanogluetal04,leeetal09}. Our own
observations demonstrate the absence of short-term \oc\ variations
with periods less than 1\,yr exceeding a couple of seconds.  We have
scrutinizingly surveyed the available data and found that a small
number of published mid-eclipse times deviate significantly from the
mean \oc\ variation, suggesting that the errors were
underestimated. Rather than including all data indiscriminately, we
excluded these outliers from our analysis on the assumption that the
absence of short-term \oc\ variations as documented by the SAAO
timings and our own data holds at all times. In general, we accepted
mid-eclipse times if they have a quoted error not exceeding
0.0001\,day. An adopted systematic error of 3.0\,s was quadratically
added to all eclipse times included in our data base. We restricted our
analysis to primary mid-eclipse times, because the times for the
secondary eclipses are significantly less well-determined and add
little to the definition of the long-term \oc\ variation. The mean
phase of the secondary eclipses agrees with $\phi\!=\!0.500$ within a
few seconds, similar to our finding for the sdB systems \nsvs\ and
\hs\ (paper II).

\begin{table}[t]
\begin{flushleft}
\caption{Parameters of the eclipsing binary \hw. }
\begin{tabular}{lll}
\hline \hline\\ 
Parameter & Value$\pm$ Error & Reference\\[1ex] 
\hline \\ 
Orbital period $P_\mathrm{\,bin}$ ~(s)    & $10084.58\pm 0.01$   & This work \\ 
Primary mass~$M_1$ (\msun)  & $0.485\pm 0.013$ & Lee et al. (2009) \\
Secondary mass~$M_2$ (\msun)  & $0.142\pm 0.004$ & Lee et al. (2009) \\
Secondary radius~$R_2$ (\rsun)  & $0.175\pm 0.026$ & Lee et al. (2009) \\
Separation $a_\mathrm{\,bin}$ (\rsun) & $0.860\pm 0.010$ & Lee et al. (2009)\\
Inclination $i_\mathrm{\,bin}$ ($^\circ$) & $80.9\pm0.1$ & Lee et al. (2009)  \\ 
Eccentricity $e_\mathrm{\,bin}$         & $<0.0003$ & This work  \\ 
Distance $d$ (pc) & $181\pm 20$ & Lee et al. (2009) \\
Visual magnitude $V$ (mag) & 10.9 &  \\[1ex]
\hline\\
\end{tabular}
\end{flushleft}
\label{tab:hw}
\vspace{-2mm}
\end{table}

Figure~\ref{fig:data} shows the \oclinl\ residuals of the data set
adopted by us relative to the linear ephemeris used by
\citet{leeetal09} in their Fig.~5 (top panel). We have included all
111 primary mid-eclipse times reported by the SAAO group, 20 from
\citet{leeetal09}, four from \citet{woodetal93} and
\citet{woodsaffer99}, three from the OEJV \citep{bratetal11}, eight
VSNET times, three BAV times, and the new mid-eclipse times of
Table~\ref{tab:monetaavso}. Our entire data base contains 176 primary
mid-eclipse times. We have corrected all times to the Solar-system
barycenter in the terrestrial system quoting them as
BJD(TT)\footnote{http://astroutils.astronomy.ohio-state.edu/time/}.

In Table~\ref{tab:hw}, we summarize the parameters of the binary \hw\
relevant to the present study. The masses and the distance are taken
from \citet{leeetal09}. The limit on the eccentricity was obtained
from our new mid-eclipse times $T_\mathrm{ecl}$ in
Table~\ref{tab:monetaavso}, which limit the amplitude caused by
apsidal motion to $\Delta
T_\mathrm{ecl}\!\simeq\!P_\mathrm{bin}\,e/\pi\!<\!1.0$\,s for a period
of the apsidal rotation
$U\!\simeq\!P_\mathrm{bin}\,(M_2/M_1)(a_\mathrm{bin}/R_2)^5/(15\,k_2)\!\simeq\!43$\,d,
with $k_2\!\simeq\!0.15$ the structure constant of the nearly fully
convective secondary star \citep{feidenetal11}.

\section{The Lee et al. model}

In an influential paper, \citet{leeetal09} interpreted the mid-eclipse
times of \hw\ available until 2008 (cycle number $E\!=\!76050$) by the
LTT effect of two planets superposed on a quadratic variation of
unspecified origin.  Figure~\ref{fig:data} shows their model curves for the quadratic
variation (dashed curve) and their final model (solid curve). Their
fit is adequate until 2008, but fails completely to reproduce our new
data. The discrepancy has reached 250\,s in 2012 and is largely due to
the rapid fall-off of the quadratic term. The continuous period decrease as
defined by the quadratic term does not exist.

The \citet{leeetal09} model faces the additional problem that the
proposed 2-planet system is secularly unstable. This result is evident
from the planetary parameters quoted by them, which imply crossing
orbits with an apoapsis of the inner planet of 4.7\,AU and a periapsis
of the outer planet of of 2.9\,AU. We have numerically integrated the
orbits with {\tt mercury6} \citep[][see below for more
  details]{1999MNRAS.304..793C} and find that a near encounter or a
collision occurs within 2000\,yr. Hence, the model of
\citet{leeetal09} is untenable in the present form.

In the remainder of the paper, we show that all observations can be
explained by the LTT effect, without taking recourse to an additional
mechanism.  A minor contribution by gravitational radiation is not
excluded, but is below our detection limit.

\begin{figure} [t]
\centering
\includegraphics[width=89mm]{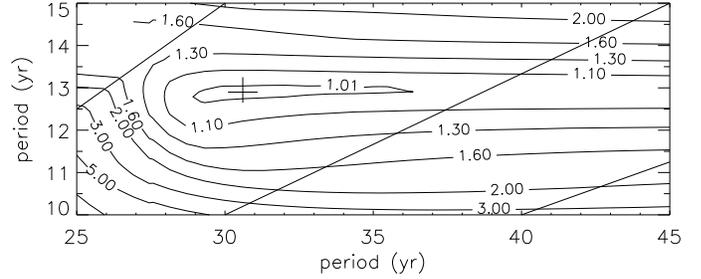}
\caption{Contour plot of the reduced $\chi^2$ normalized to unity for
  the best fit using a generalized Lomb-Scargle periodogram for
  a two-companion fit to the eclipse-time variations of \hw. The solid
  lines indicate 2:1, 3:1 and 4:1 resonant orbits, the cross the best
  fit. Solutions with an inner planet of about 12.7 year orbital
  period provide good fits for a wide range of orbital periods of the
  outer companion.}
\label{fig:GLS}
\end{figure}

\section{Search method}

We first scanned the two-companion parameter space for orbital periods
below 60 years, using a generalized Lomb-Scargle periodogram
supplemented by refined local searches. In a second step, we searched
for improved solutions using a Levenberg-Marquardt (LM) minimization
algorithm … allowing us to detect the true local minima near the start
parameters. Finally, we tested the secular stability of all possible
solutions using the hybrid symplectic algorithm implemented in the
{\tt mercury6} package \citep{1999MNRAS.304..793C}. 

Fitting Keplerian orbits to the set of mid-eclipse times involves a
simultaneous fit for the ephemeris of the binary star (two parameters)
and the orbits of the companions (five parameters each). The orbital
inclination $i$ of a companion remains undetermined. For two or more
companions, a direct search of the parameter space is computationally
elaborate and time-consuming because of the large number of parameters
involved. We adopted, therefore, the approach of
\cite{2009A&A...496..577Z} and developed a variant of the generalized
Lomb-Scargle (GLS) periodogram \citep{1976Ap&SS..39..447L,
  1982ApJ...263..835S, 1986ApJ...302..757H}. The mid-eclipse time of
eclipse number $n$ including the light-travel time of planets
$k\!=\!1\ldots N_\mathrm{p}$ is expressed as
\begin{equation}
  T(n)=T_0+n\,P_\mathrm{bin} + \sum_{k=1}^{N_\mathrm{p}}K_{k}\frac{1-e_{k}^2}
  {1\!+\!e_{k}\cos\nu_{k}(n)}\sin\left(\nu_{k}(n)+\omega_{k}\right),
\label{EQtiming}
\end{equation}
where $e_{k}$ is the eccentricity of planet $k$,
$K_{k}\!=\!a_{\mathrm{bin},k}\sin\,i_{k}/c$ the amplitude of the
LTT effect, with $a_{\mathrm{bin},k}$ the semi-major axis of the orbit of the
center of gravity of the binary around the common center of mass and
$i_k$ the inclination, $\omega_{k}$ is the argument of
periastron measured from the ascending node in the plane of the sky,
and $\nu_{k(n)}$ the true anomaly at time $t=T_0+n\,P_\mathrm{bin}$
\citep{kopal59}.  Eq.~\ref{EQtiming} can be transformed into a linear
equation for the coefficients $A_{k}$, $B_{k}$, and $C$, using the
eccentric anomaly $E_{k}(n)$ of planet $k$ at the time of
eclipse $n$,
\begin{equation}
T(n) = n\,P_\mathrm{bin}+C+\sum_{k=1}^{N_p}\big (A_k\cos E_k(n)+B_k\sin E_k(n)\big )
\end{equation}
with \\[-4ex]
\begin{displaymath}
A_{k}\!=\!-K_{k}\sin\omega_{k}, ~B_{k}\!=\!K_{k}\!\sqrt{1\!-\!e_{k}^2}\cos\omega_{k},
~C\!=\!T_0\!+\!\!\sum_{k=1}^{N_\mathrm{p}}K_{k}e_{k}\sin\omega_{k}. 
\end{displaymath}
The best-fit values of $T_0$, $ P_\mathrm {bin}$, $K_{k}$, and
$\omega_{k}$ were derived by linear regression for given values
of $e_{k}$, orbital period $P_{k}$ and time
$t_{k}$ of periastron passage. Calculating a grid in these
quantities reduces the number of free parameters for each grid point
from $2+5\,N_\mathrm{p}$ to $3\,N_\mathrm{p}$.

In a second step, we searched for improved solutions near the best
solution from GLS scan. We employed the Levenberg-Marquardt (LM)
minimization algorithm \citep{2009ASPC..411..251M}, a non-linear
least-squares fitting routine implemented in {\tt mpfit} of {\tt IDL}.
We used the GLS-parameters as start values and generated 500
variations of them, using the diagonal elements of the {\tt
  mpfit}-derived covariance matrix as uncertainties $\sigma$. The
start parameters were varied randomly within $\pm 10~\sigma$ around
the original start values, allowing us to detect the true local
minimum near the start parameters.

\begin{figure}[t]
%
\includegraphics[height=89mm,angle=-90,clip]{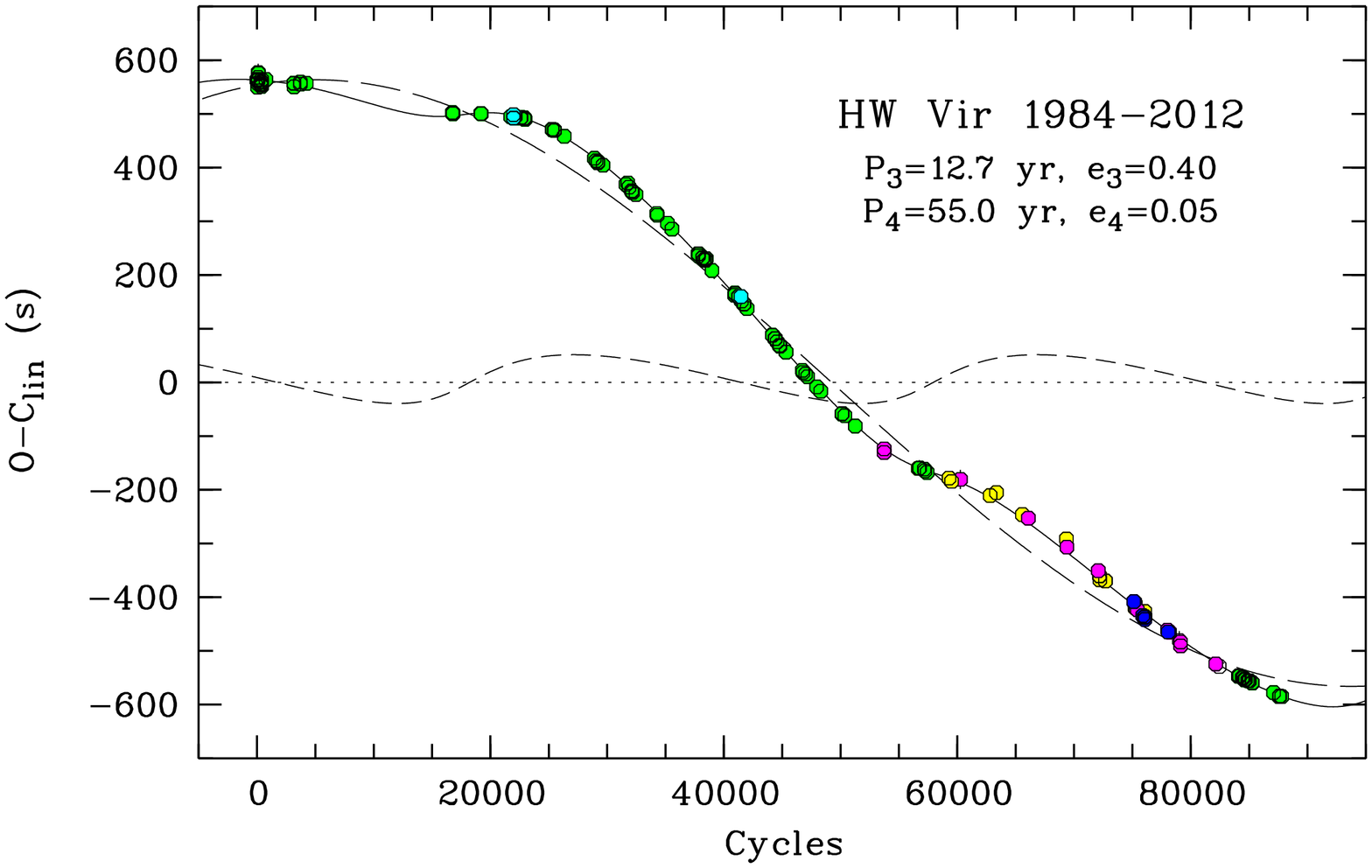}
\includegraphics[height=89mm,angle=-90,clip]{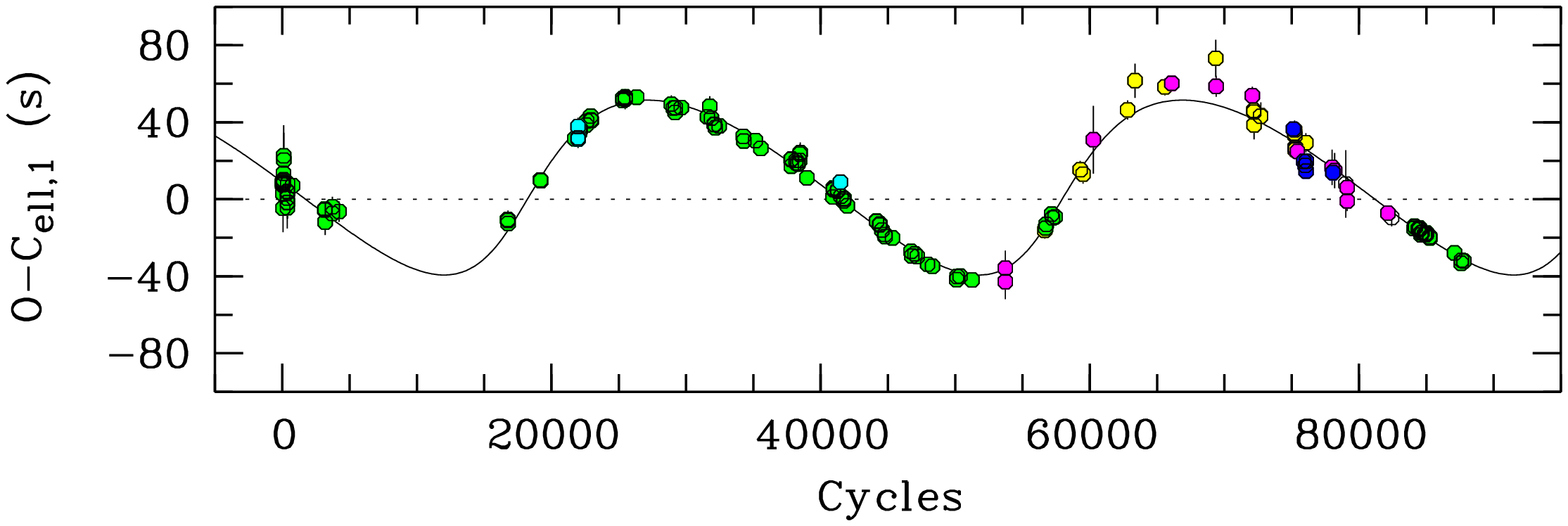}
\includegraphics[height=89mm,angle=-90,clip]{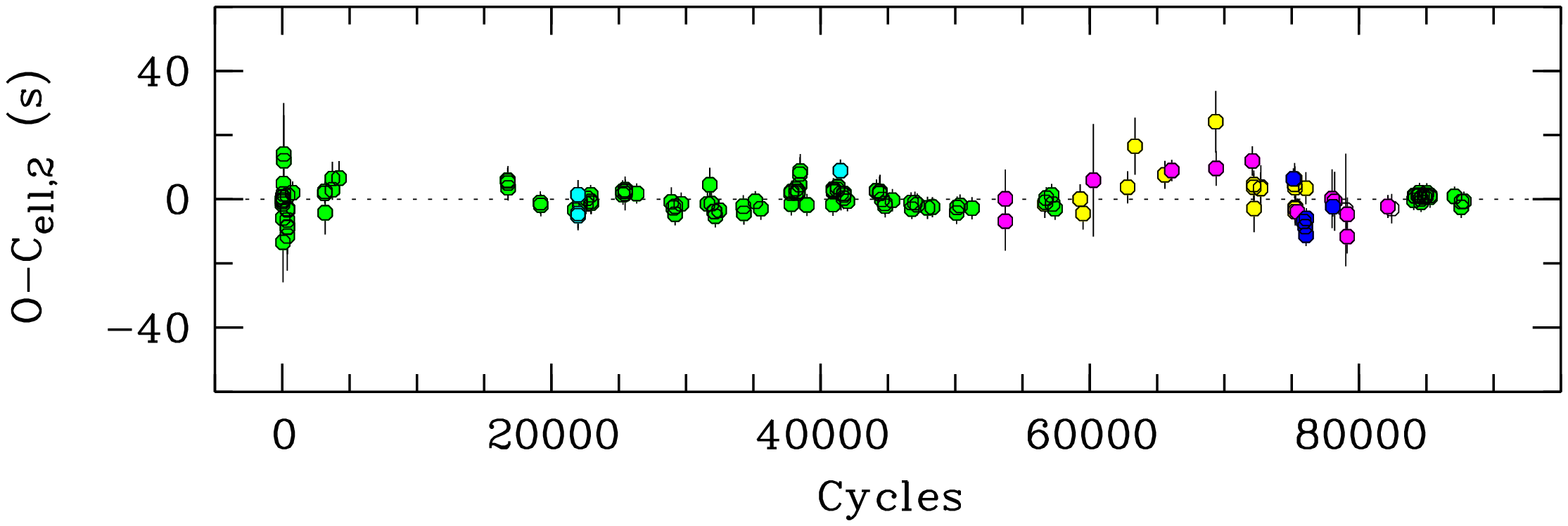}
\caption[chart]{Fit of two Keplerian orbits to the eclipse-time
  variations of \hw. \emph{Top:} Data of Fig.~\ref{fig:data} relative
  to the linear ephemeris of Eq.~\ref{eq:ephem}. The curves denote the
  model LTT effect (solid) and the contributions by the outer
  companion (long dashes) and the inner planet (short
  dashes). \emph{Center:} Data with the contribution by the outer
  companion subtracted and model for the inner planet (solid
  curve). \emph{Bottom:} Residuals after the subtraction of the
  contributions by both companions.}
\label{fig:fit}
\end{figure}

Finally, we tested the secular stability of the solutions, using the
hybrid symplectic integrator in {\tt mercury6}
\citep{1999MNRAS.304..793C}, which allows one to evolve planetary
systems with high precision over long times very efficiently. We used
constant time steps of 35\,d. Test runs demonstrated that this choice
is adequate and smaller steps do not change the results.  This
provision is not adequate for the treatment of a close encounter, but
such incidence should not occur in the successful models and if it
does, the calculation is stopped and the model termed 'unstable'.
The accuracy parameter in the code was set to $10^{-16}$, leading to a
fractional change in the energy and momentum of the triple (binary and
two companions) of typically $10^{-5}-10^{-4}$ and $10^{-9}-10^{-8}$,
respectively. The quoted changes are valid for an integration time of
$10^7$\,yr and the larger values obtain for longer periods $P_4$ and,
hence, larger masses $M_4$. For simplicity, the central binary was
treated as a single object with a mass equal to the sum of the
component masses. This simplification is justified given the short
binary orbital period of 2.8\,h. The gravitational field at the
position of a distant companion can be represented as the sum of the
constant field created by the combined mass of the binary components
and a gravitational wave, emanating from the revolving binary with
periods of 2.8\,h for the fundamental and 1.4\,h for the first
harmonic. The relative strength of the wave field is $\ten{4}{-8}$ and
closely averages to zero over the 300 or 600 periods that occur in a
time step of 35\,d. The retroaction of the companion tends to excite
an eccentricity in the binary, but with a relative strength of
$10^{-8}$ this effect is also entirely negligible. We used the masses
of the binary components as given by \citet{leeetal09} and quoted in
Table~\ref{tab:hw}. The orbital evolution of the companions was
followed until instability occurred, or at least for $10^7$\,yr and up
to $10^8$\,yr for some models.

\section{A stable two-companion model for \hw}

\begin{table}[b]
\begin{flushleft}
  \caption{Parameters of the two-companion model of Fig.\,\ref{fig:fit}
    for \hw. A colon indicates uncertain values.}
\label{tab:companions}
\begin{tabular}{l@{\hspace{0mm}}c@{\hspace{3mm}}c}
\hline \hline\\ 
Parameter & Inner companion & Outer companion\\[1ex] 
          & $\#$\,3 & $\#$\,4 \\[1ex] 
\hline \\ 
Orbital period $P$ ~(yr)    & $12.7\pm0.2$   & $55~~(\mathrm{fixed})$ \\ 
Eccentricity $e$            & $0.40\pm 0.10$ & $0.05\,:$  \\ 
Semi-major axis $a$ ~(AU)   & $4.69\pm0.06$  & $12.8\pm 0.2$       \\ 
Amplitude $K$ ~(s)          & $49\pm 3$      & $563\pm 200$   \\ 
Mass ~~$M$\,sin\,$i$ ~(\mjup) & $14.3\pm 1.0$  & $65\pm 15$ \\ 
Argument of periastron ~$\omega$ ~($^\circ$) & $-18\pm 10$   & 0 :    \\
Periastron passage ~(JD)    & $2\,452\,401~$   & $2\,461\,677$ :    \\
Periastron passage ~(Cycle) & $57\,150~$   & $136\,619$ :    \\[1ex]
\hline\\
\end{tabular}
\end{flushleft}
\end{table}

The data shown in Fig.~\ref{fig:data} suggest the presence of a period
near 40000 cycles or 13\,yr superposed on a variation with a longer
period. Using the GLS periodogram, we scanned the parameter space for
two companions with orbital periods of $P_3=10-15$\,yr and
$P_4/P_3\!\ge\!1.85$. Figure~\ref{fig:GLS} shows a contour plot of the
reduced $\chi^2$ normalized to unity for the best fit. For a better
presentation, only the range $P_4\!\le\!45$\,yr is displayed; no new
features appear for larger $P_4$. A ridge of low \chisq\ confirms
$P_3\!\simeq\!12.5-12.9$\,yr for all $P_4\!\ga\!28$\,yr, including the
possible mean-motion resonances with $P_4\!:\!P_3$ = 5\,:\,2, 3\,:\,1,
7\,:\,2, 4\,:\,1, and 5\,:\,1. None of the resonant solutions,
however, is preferred over non-resonant ones. For the best fit near
$P_4\!=\!30$\,yr , the inner object has a mass close to the boundary
between planets and brown dwarfs, the outer one is a brown dwarf, with
a mass increasing with orbital period up to the stellar mass limit for
$P_4\!\simeq\!70$\,yr.

\begin{figure*}[t]
\includegraphics[height=91mm,angle=-90,clip]{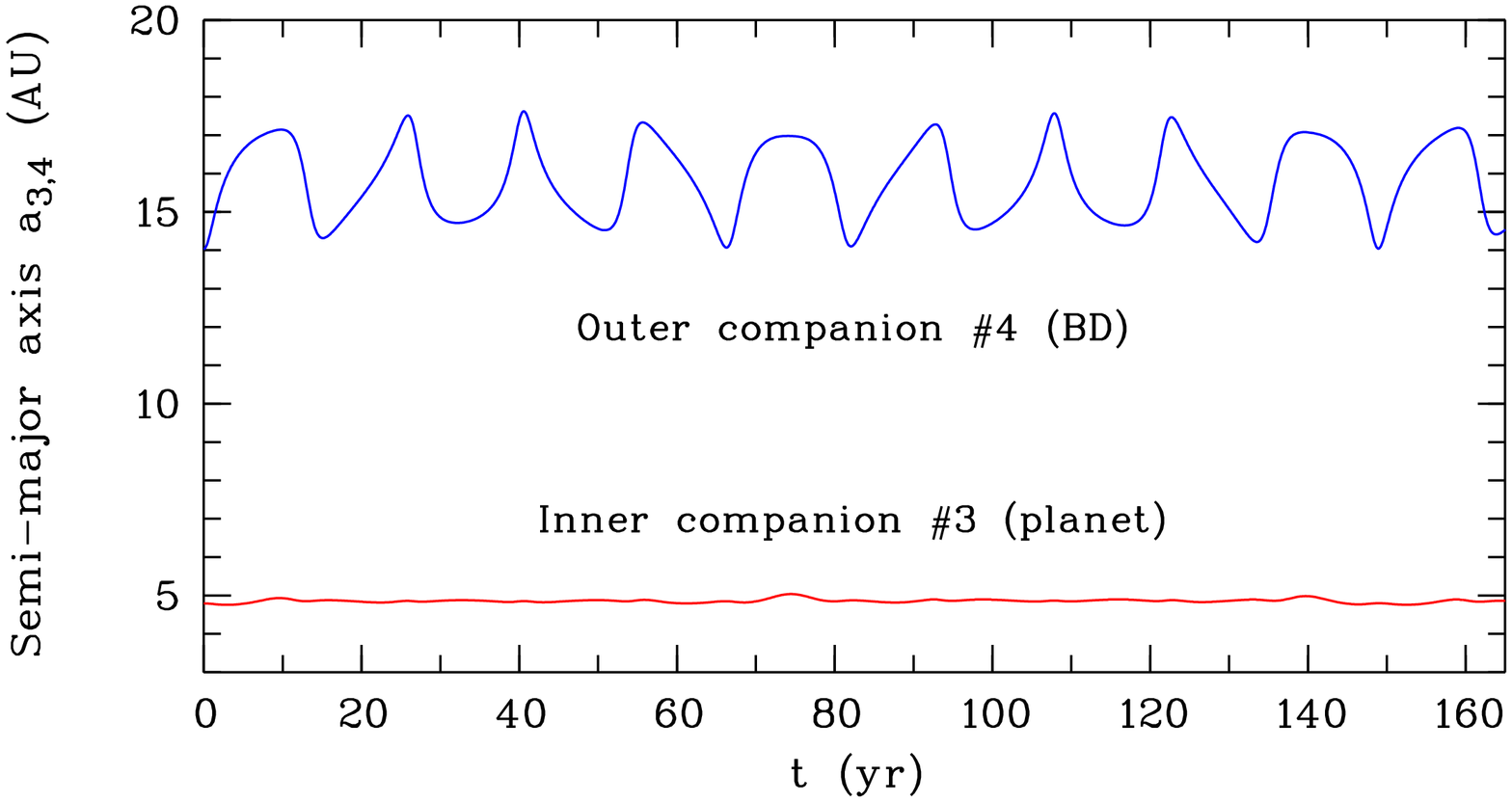}
\hfill
\includegraphics[height=91mm,angle=-90,clip]{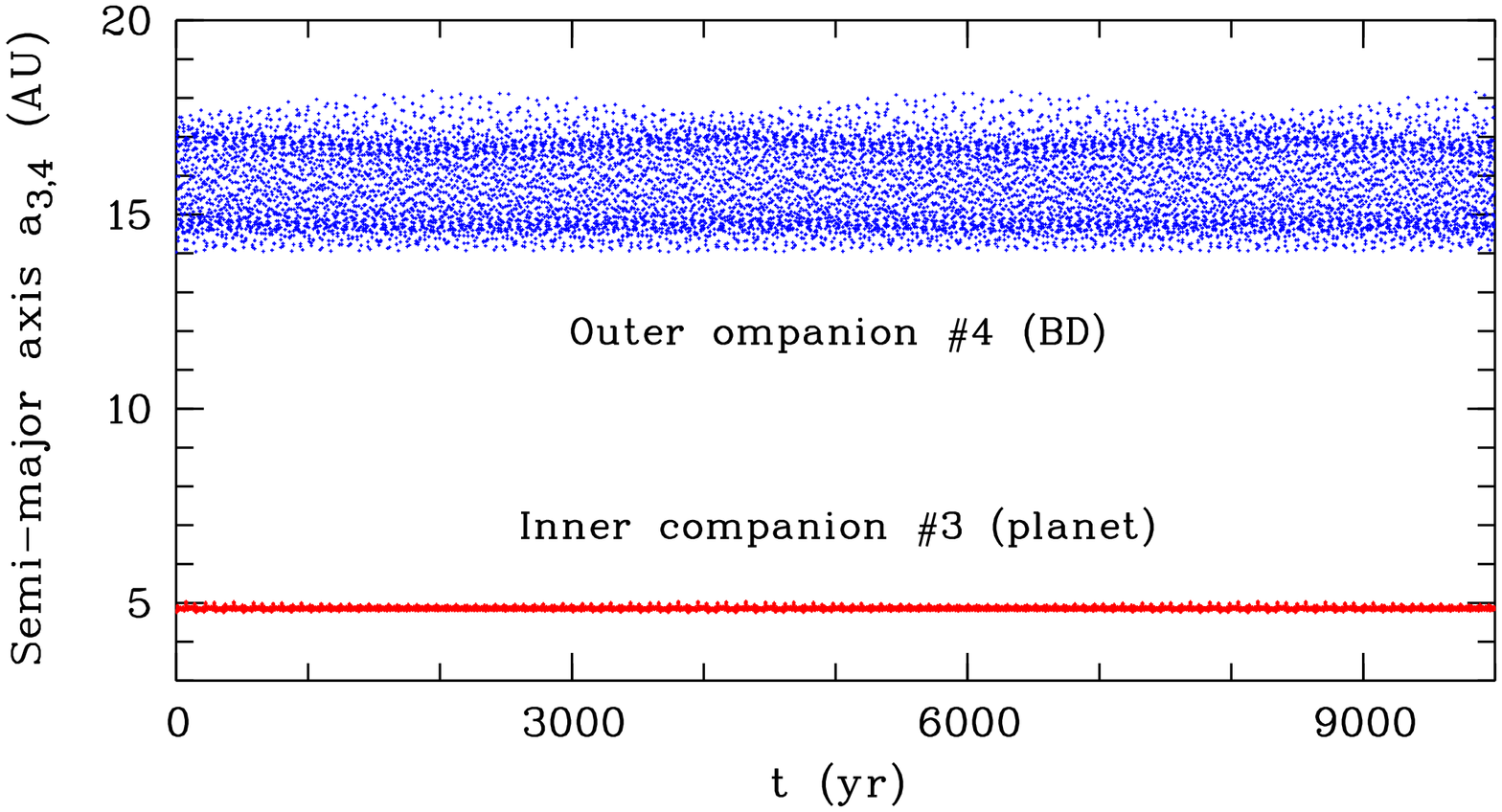}

\includegraphics[height=91mm,angle=-90,clip]{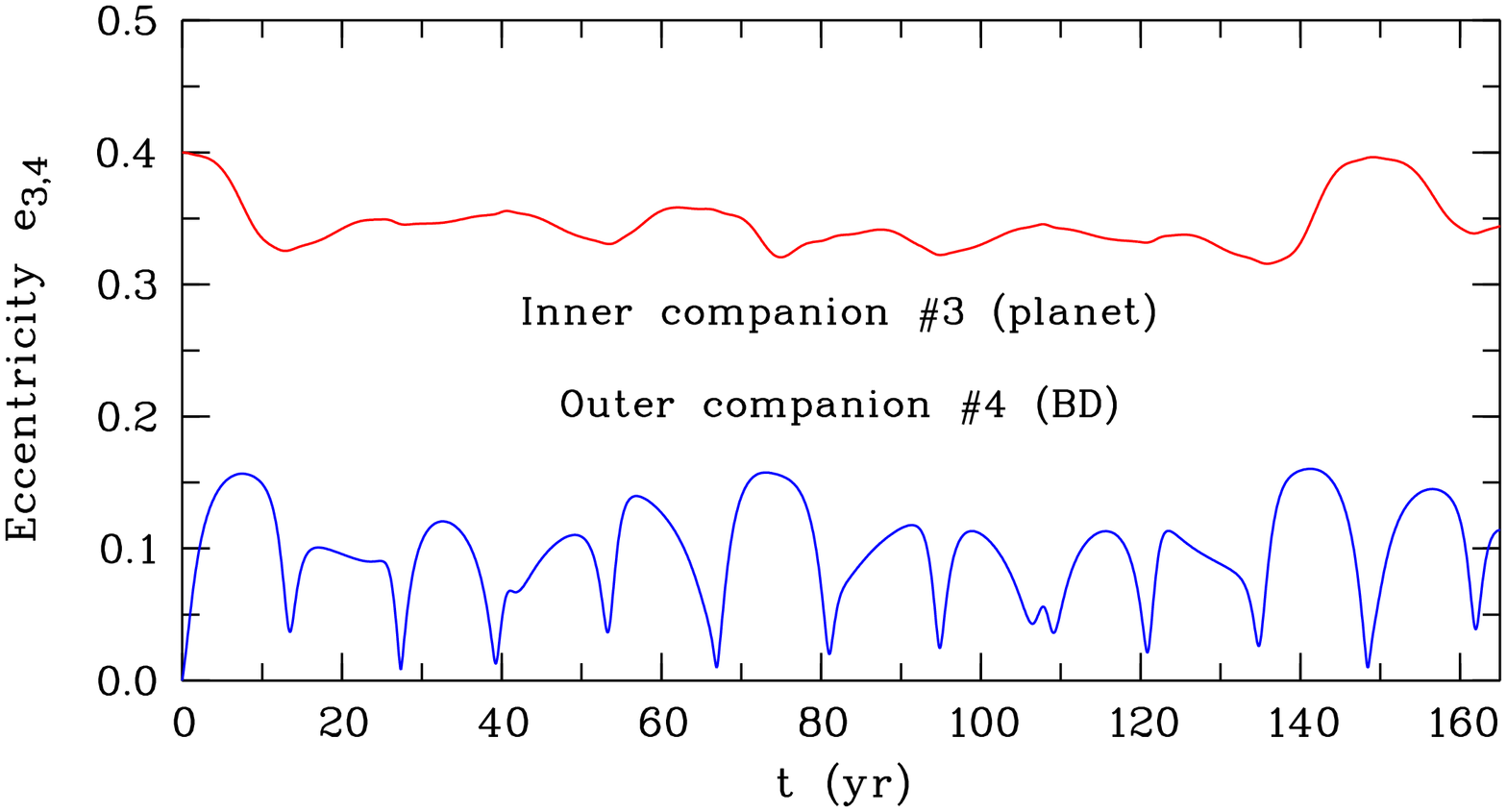}
\hfill
\includegraphics[height=91mm,angle=-90,clip]{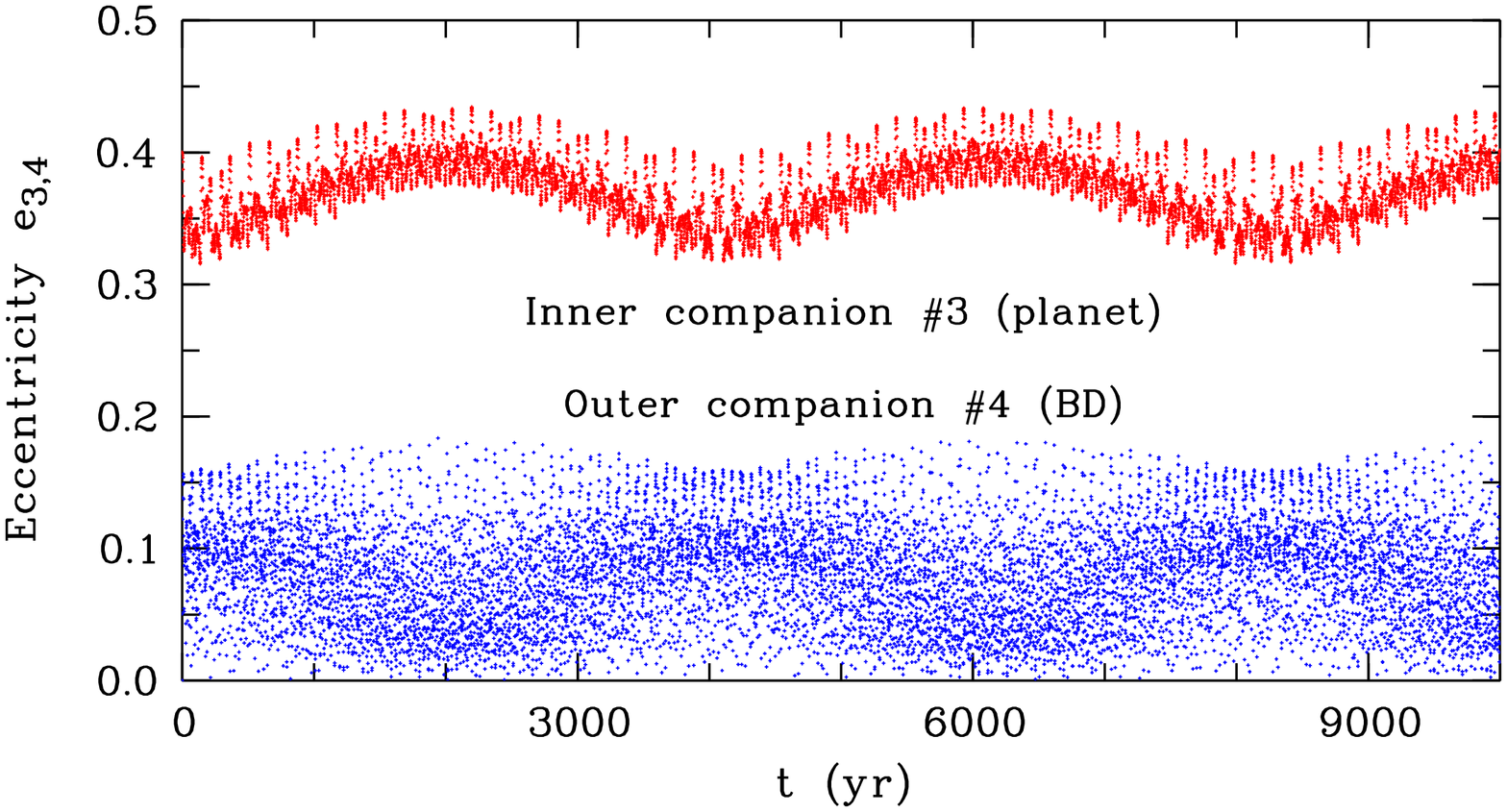}
\caption[chart]{Temporal variation of the semi-major axes and the
  eccentricities of the osculating orbits of the companions to \hw\ as
  calculated with the Mercury6 code (see text). The system is stable
  for more than $10^8$\,yr (see text). Color coding is the same for
  all panels.}
\label{fig:model}
\end{figure*}

With the LM optimization algorithm, we found improved two-companion
solutions along the ridge of low \chisq\ in
Fig.~\ref{fig:GLS}. Formally, the best fit is now obtained near a
period ratio of 4\,:\,1 with periods $P_3\!=\!12.7$\,yr and $P_4$
around $50$\,yr. The fit deteriorates for $P_4\!>\!70$\,yr. The
eccentricity of the inner companion is rather well defined with
$e_3\!=\!0.4\pm0.1$, that of the outer one is not, although small
finite values fit better than $e_4\!=\!0$. Since the data cover only
part of the orbital period of the outer companion, the deduced
amplitude $K_4$ of the LTT effect depends on the adopted values of
$P_4$ and $e_4$. Aided by the stability calculations described below,
we chose of $P_4\!=\!55$\,yr and $e_4\!=\!0.05$.  Figure~\ref{fig:fit}
(top panel) shows the best fit and Table~\ref{tab:companions}
summarizes the fit results with our best estimates of the errors.  The
center panel repeats the data with the dominant contribution from the
outer companion subtracted ($O\!-\!C_\mathrm{ell,1}$) and the bottom
panel shows the residuals after subtraction of both contributions
($O\!-\!C_\mathrm{ell,2}$). Small systematic residuals exist, notably
around cycle number 70000, where the data quality is not the best,
however.  The formal best fit leads to a symmetric configuration with
$\Delta \omega\!=\!\omega_4\!-\!\omega_3$ near zero, but with a large
uncertainty in $\omega_4$, because of the limited data coverage. The
general decrease of \oclin\ over the 1984--2012 time interval reflects
the fact that the observations cover only that part of the orbit, when
the outer companion is receding from the observer and the binary is
approaching. The corresponding LTT amplitude suggests that the outer
companion is probably a brown dwarf with a mass of roughly 65\,\mjup\
or a star of very low mass. For periods between 40 and 70\,yr, its
mass would be in the range of 30 to 120\,\mjup. The fit yields an
underlying linear ephemeris of the binary
\begin{equation}
T_\mathrm{ecl} = \mathrm{BJD(TT)}\,2445730.5497(72) + 0.11671969(15) E,
\label{eq:ephem}
\end{equation}
where the quoted correlated uncertainties of the epoch and the binary
period $P_\mathrm{bin}$ refer to the quoted range of $P_4$.

Using the {\tt mercury6}~code, we followed the orbital evolution of
solutions along the low-\chisq\ ridge in Fig.~\ref{fig:GLS}
numerically, using combinations of periods $P_3$ and $P_4$ and masses
$M_3$ and $M_4$ that match the observations. We note that the actual
positions of the two companions in space, their orbital velocities,
and true anomalies at a given time can not be obtained with sufficient
accuracy from the observations. We employed, therefore, standardized
start parameters and searched for stable models that reproduce the
observations for a substantial percentage of the time. The
calculations were started in either the symmetric or the antisymmetric
configuration with both companions at periastron, having Keplerian
velocities. The orbits were taken to be coplanar with the binary orbit
($i\!=\!80.9^\circ$, Table~\ref{tab:hw}), with the start
eccentricities as free parameters. Orbital evolution changes the
elements quickly and periodically varying eccentricities develop also
for models that start with circular orbits.

We dubbed a model 'stable' if it persisted without close encounter for
more that $10^7$\,yr. Such models are found for periods $P_4$ as short
as 35\,yr, but for most start configurations at this period
instability incurred, sometimes only after $10^6$\,yr. The fraction of
stable models that mimic the observations becomes substantial only
near 45\,yr and at 55\,yr or longer. A low fraction of stable models
near 37\,yr and, less so, near 50\,yr may imply a higher probability
of instability near the 3:1 and 4:1 mean-motion resonances.  
The general preference for longer periods $P_4$ is expected, because
the observations fix $P_3$ between 12.5 and 12.9\,yr and a more
distant outer companion increases the phase space for solutions with
larger eccentricities. At $P_4\!\ge\!55$\,yr, stable models were found
for eccentricities $e_3$ as large as 0.45. The properties of stable
models between $P_4\!=\!45$\,yr and 55\,yr differ in that the former
prefer anti-symmetric configurations, whereas the latter all spend a
substantial fraction of the time near a symmetric configuration.
Taken at face value, our observational best fit of $\Delta \omega$
near zero favors a model with $P_4\!>\!50$\,yr. Our preferred model
has $P_4\!=\!55$\,yr, values of $P_3$, $e_3$, $M_3$, and $ M_4$ as
obtained from the fit to the data (Fig.~\ref{fig:fit},
Table~\ref{tab:companions}), and orbits coplanar with the binary.

Figure~\ref{fig:model} shows the evolution of the semi-major axes and
eccentricities for the preferred model. The left panels illustrate the
behavior for the first 165\,yr (three orbital periods of the outer
companion), the right panels show the evolution over 10\,000\,yr. The
system is stable for $10^8$\,yr. The elements of the osculating orbits
vary periodically on time scales up to 4000\,yr (right panels). The
eccentricities $e_3$ and $e_4$ range from 0.32 to 0.43 and from zero
to 0.18, respectively. Over short time intervals, the modulation at
the synodic period of the inner planet
$P_\mathrm{syn}\!=\!(P_3^{-1}-P_4^{-1})^{-1}\!\simeq\!16$\,yr is dominant (left
panels). The system spends a substantial fraction of the time near the
observed eccentricities, $e_3\!\simeq\!0.4$ and $e_4\!\simeq\!0.05$,
and assumes a symmetric configuration most frequently near maximum
$e_3$ in the 4000\,yr cycle (lower right panel), suggesting that the
observational fit of Fig.~\ref{fig:fit} and Table~3 represents a
snapshot of the variable constitution of the \hw\ system. While the
agreement seems satisfactory, we add the caveat that our simulations
have so far only provided a first glimpse at the stability landscape,
which remains to be explored in more detail.

The perturbation of the orbits by the mutual interaction of the two
companions leads to non-Keplerian shifts of the observed eclipse
times, which are not included in the fit shown in Fig.~\ref{fig:fit}.
Our simulations show that such shifts are not negligible and affect
the derived values of the LTT amplitude, the period, the eccentricity,
and the argument of periastron of both companions. In particular, the
LTT amplitude of the inner companion can vary by more than 10\,s
between subsequent orbits. Such orbit-to-orbit variations are not
correctly interpreted by our present fitting routines. The observed
residuals $O\!-\!C_\mathrm{ell,2}$ after subtraction of the
contributions from the two Keplerian contributions may show a
non-random structure, possibly representing the signature of the
perturbed orbits (Fig.~\ref{fig:fit}, bottom panel), but the quality
of the data is lowest between cycle numbers 60000 and 80000, where the
excursion appears largest. Clearly, a unique identification of
non-Keplerian eclipse-time variations in PCEB would provide a strong
support for the planetary model.

We have alternatively considered that the residuals in Fig.~\ref{fig:fit} might
indicate the presence of a third inner planet. While the added free
parameters improve the fit, this possibility can be excluded, because
such object is quickly expelled.

\section{Conclusion}

We have investigated the plausibility of the planetary model for
HW\,Vir, using additional data and an exhaustive search for and
dynamical tests of two-companion solutions. The original model by Lee et
al. (2009) is clearly ruled out, both from the current direction of
the observed \oc\ variations as well as the fact that their solution
is dynamically unstable. We have presented a qualitatively different
solution that involves two companions in secularly stable orbits about
the binary \hw. The inner companion has an orbital period
$P_3\!\simeq\!12.7\pm0.2$\,yr and a mass $M_3\!\simeq\!14.3\pm
1.0$\,\mjup\ for an inclination $i_3$ identical to that of the binary
($i\!=\!80.9^\circ$, see Table~\ref{tab:hw}). For other inclinations,
the mass varies as 1/sin\,$i_3$.  The outer more massive companion is
a brown dwarf or a low-mass star with a mass around
$M_4\,\mathrm{sin}\,i_4\!\simeq\!30$ to 120\,\mjup, where $i_4$ is the
unknown inclination of its orbit. Its parameters are still uncertain,
because the data cover only about half an orbital period.  While the
suggested pair of companions in coplanar orbits is secularly stable,
we caution that the stability landscape has not been thoroughly
explored and its systematic structure is still eluding us. The system
is strongly interacting with evidence for non-Keplerian orbits, whose
signature may even be visible in the data.

Our model of HW\,Vir provides an attractive possibility of explaining
the entire observed orbital-period variations of \hw\ by the LTT
effect, without taking recourse to any additional unexplained
process. It provides positive evidence of the planetary model for
\hw\ and strengthens the case for close binaries in general. It also
calls for a re-examination of other systems in the light of our
result. However, the case of \hw\ also shows how difficult it may be to
find such a solution, given both the very long time-series required to
detect clearly periodic planetary signals, the requirement of fitting
all residuals without invoking other arbitrary mechanisms, and the
limits of fitting individual static Keplerian orbits to what must
obviously be very dynamic systems.

\begin{acknowledgements}
  We thank the anonymous referee for helpful comments that improved
  the presentation. This work is based in part on data obtained with
  the MOnitoring NEtwork of Telescopes (MONET), funded by the Alfried
  Krupp von Bohlen und Halbach Foundation, Essen, and operated by the
  Georg-August-Universit\"at G\"ottingen, the McDonald Observatory of
  the University of Texas at Austin, and the South African
  Astronomical Observatory. We acknowledge with thanks the variable
  star observations from the AAVSO International Database contributed
  by observers worldwide and used in this research. The work also
  benefitted from the use of data made available by the Variable Star
  Network, World Center for Transient Object Astronomy and Variable
  Stars. Finally, we thank Tim-Oliver Husser for measuring the eclipse
  on 21 January 2011.

\end{acknowledgements}

\bibliographystyle{aa}

\begin{thebibliography}{29}
\expandafter\ifx\csname natexlab\endcsname\relax\def\natexlab#1{#1}\fi

\bibitem[Agerer \& Hubscher(2000)]{agerer00} Agerer, F., \& H\"ubscher,
  J.\ 2000, IBVS, 4912, 1
\bibitem[Agerer \& H\"ubscher(2002)]{agerer02} Agerer, F., \&
  Hubscher, J.\ 2002, IBVS, 5296, 1
\bibitem[Agerer \& Hubscher(2003)]{agerer03} Agerer, F., \& H\"ubscher,
  J.\ 2003, IBVS, 5484, 1
\bibitem[Applegate(1992)]{applegate92} Applegate, J.~H.\ 1992, 
\apj, 385, 621 
\bibitem[Beuermann et al.(2010)]{beuermannetal10} Beuermann, K.,
  Hessman, F. V., Dreizler, S. et al., 2010, \aap, 521, L60
\bibitem[Beuermann et al.(2011)]{beuermannetal11} Beuermann, K.,
  Buhlmann, J., Diese, J., et al.\ 2011, \aap, 526, A53
\bibitem[Beuermann et al.(2012)]{beuermannetal12} Beuermann, K.,
  Breitenstein, P., D\d{e}bski, B.~D., et al.\ 2012, \aap, 540, A8 (paper II)
\bibitem[Br{\'a}t et al.(2008)]{bratetal08} Br{\'a}t, L., {\v 
S}melcer, L., Ku{\`e}{\'a}kov{\'a}, H., et al.\ 2008, OEJVS, 94, 1 
\bibitem[Br{\'a}t et al.(2009)]{bratetal09} Br{\'a}t, L., Trnka, J., 
Lehky, M., et al.\ 2009, OEJVS, 107, 1 
\bibitem[Br{\'a}t et al.(2011)]{bratetal11} Br{\'a}t, L., Trnka, J.,
{\v S}melcer, L., et al.\ 2011, OEJVS, 137, 1
\bibitem[Brinkworth et al.(2006)]{brinkworthetal06} Brinkworth, C.~S., 
Marsh, T.~R., Dhillon, V.~S., \& Knigge, C.\ 2006, \mnras, 365, 287 
\bibitem[{\c C}akirli \& Devlen(1999)]{cakirlidevlen99} {\c C}akirli ,
  {\"O}., \& Devlen, A.\ 1999, \aaps, 136, 27
\bibitem[Chambers(1999)]{1999MNRAS.304..793C} Chambers, J.~E.\ 1999,
  \mnras, 304, 793
\bibitem[Doyle et al.(2011)]{kepler16} Doyle, L.~R., Carter, 
J.~A., Fabrycky, D.~C., et al.\ 2011, Science, 333, 1602 
\bibitem[Feiden et al.(2011)]{feidenetal11} Feiden, G.~A., Chaboyer, 
B., \& Dotter, A.\ 2011, \apjl, 740, L25 
\bibitem[G\"urol \& Selan(1994)]{gurolselan94} G\"urol, B., \& Selan,
  S.\ 1994, IBVS, 4109, 1
\bibitem[Henden(2010)]{aavso} Henden, A.~A.\ 2010, Observations from
  the AAVSO International Database, private communication.
\bibitem[Hinse et al.(2012)]{hinseetal12} Hinse, T.~C., Lee, J.~W., 
Go{\'z}dziewski, K., et al.\ 2012, \mnras, 420, 3609 
\bibitem[Horne \& Baliunas(1986)]{1986ApJ...302..757H}Horne, J.~H. \&
Baliunas, S.~L.\ 1986, \apj, 302, 757
\bibitem[Horner et al.(2011)]{horneretal11} Horner, J., Marshall, 
J.~P., Wittenmyer, R.~A., \& Tinney, C.~G.\ 2011, \mnras, 416, L11 
\bibitem[{\.I}bano{\v g}lu et al.(2004)]{ibanogluetal04} {\.I}bano{\v
    g}lu, C., {\c C}ak{\i}rl{\i}, {\"O}., Ta{\c s}, G., \& Evren, S.\
  2004, \aap, 414, 1043
\bibitem[Kato et al.(2004)]{vsnet} Kato, T., Uemura, M., Ishioka, R.,
  et al.\ 2004, \pasj, 56, 1
\bibitem[Kilkenny et al.(1991)]{kilkennyetal91} Kilkenny, D.,
  Harrop-Allin, M., \& Marang, F.\ 1991, IBVS, 3569, 1
\bibitem[Kilkenny et al.(1994)]{kilkennyetal94} Kilkenny, D.,
  Marang, F., \& Menzies, J.~W.\ 1994, \mnras, 267, 535
\bibitem[Kilkenny et al.(2000)]{kilkennyetal00} Kilkenny, D., Keuris, S.,
  Marang, F., et al.\ 2000, The Observatory, 120, 48
\bibitem[Kilkenny et al.(2003)]{kilkennyetal03} Kilkenny, D., van Wyk, F.,
  \& Marang, F.\ 2003, The Observatory, 123, 31
\bibitem[Kiss et al.(2000)]{kissetal00} Kiss, L.~L., Cs{\'a}k, B.,
  Szatm{\'a}ry, K., Fur{\'e}sz, G., \& Szil{\'a}di, K.\ 2000, \aap,
  364, 199
\bibitem[Kopal(1959)]{kopal59} Kopal, Z.\ 1959, ``Close Binary
  Systems'', Chapman \& Hall, 1959, p. 109ff
\bibitem[Lee et al.(2009)]{leeetal09} Lee, J. W., Kim, S.-L., Kim,
  C.-H. et al. 2009, AJ 137, 3181
\bibitem[Lomb(1976)]{1976Ap&SS..39..447L} Lomb, N.~R.\ 1976, \apss,
  39, 447
\bibitem[Lovis et al.(2011)]{lovisetal11} Lovis, C., S{\'e}gransan,
  D., Mayor, M., et al.\ 2011, \aap, 528, A112
\bibitem[Marang \& Kilkenny(1989)]{marangkilkenny89} Marang, F., \&
  Kilkenny, D.\ 1989, IBVS, 3390, 1
\bibitem[Markwardt(2009)]{2009ASPC..411..251M} Markwardt, C.~B.\ 2009,
  ASP Conf. Ser., 411, 251
\bibitem[Nather \& Robinson(1974)]{natherrobinson74} Nather, R.~E., \&
  Robinson, E.~L.\ 1974, \apj, 190, 637
\bibitem[Parsons et al.(2010)]{parsonsetal10} Parsons, S.~G., Marsh, 
T.~R., Copperwheat, C.~M. et al.\ 2010b, \mnras, 407, 2362
\bibitem[Potter et al.(2011)]{potteretal11} Potter, S.~B., 
Romero-Colmenero, E., Ramsay, G., et al.\ 2011, \mnras, 416, 2202 
\bibitem[Qian et al.(2009)]{qian-hs0705} Qian, S- B., Zhu, L. Y., Zola,
  S. et al., 2009, ApJ 695, L163
\bibitem[Qian et al.(2010a)]{qian-hs2231} Qian, S.-B., Zhu, L.-Y.,
  Liu, L., et al.\ 2010a, \apss, 329, 113
\bibitem[Qian et al.(2010b)]{qian-dpleo} Qian, S.-B., Dai, Z.-B.,Liao,
  W.-P. et al.\ 2010b, \apjl, 708, L66
\bibitem[Qian et al.(2011)]{qian-huaqr} Qian, S.-B., Liu, L., 
Liao, W.-P., et al.\ 2011, \mnras, 414, L16 
\bibitem[Qian et al.(2012a)]{qian-rrcae} Qian, S.-B., Liu, L., Zhu, 
L.-Y., et al.\ 2012a, \mnras, 422, L24 
\bibitem[Qian et al.(2012b)]{qian-nyvir} Qian, S.-B., Zhu, L.-Y., 
Dai, Z.-B., et al.\ 2012b, \apjl, 745, L23 
\bibitem[Scargle(1982)]{1982ApJ...263..835S}Scargle, J.~D.\ 1982, \apj, 263, 835
\bibitem[Todoran(1972)]{todoran72} Todoran, I.\ 1972, \apss, 15, 229 
\bibitem[Watson \& Marsh(2010)]{watsonmarsh10} Watson, C.~A., \&
  Marsh, T.~R.\ 2010, \mnras, 405, 2037
\bibitem[Welsh et al.(2012)]{kepler34b35b} Welsh, W.~F., Orosz, 
J.~A., Carter, J.~A., et al.\ 2012, \nat, 481, 475 
\bibitem[Wittenmyer et al.(2011)]{wittenmyeretal11} Wittenmyer, R.~A.,
  Horner, J.~A., Marshall, J.~P., Butters, O.~W., \& Tinney, C.~G.\
  2011, \mnras, 1920
\bibitem[Wood et al.(1993)]{woodetal93} Wood, J.~H., Zhang,
  E.-H., \& Robinson, E.~L.\ 1993, \mnras, 261, 103
\bibitem[Wood 
\& Saffer(1999)]{woodsaffer99} Wood, J.~H., \& Saffer, R.\ 1999, \mnras, 305, 820 
\bibitem[Zechmeister \& K{\"u}rster(2009)]{2009A&A...496..577Z} Zechmeister,
  M., \&  K{\"u}rster, M.\ 2009, \aap, 496, 577

\end{thebibliography}

\end{document}